\pdfoutput=1

\documentclass[11pt,twoside,a4paper,cmspaper,final,collab]{cms-tdr}
\def\svnVersion{40826:40833M}\def\cmsCernNo{2011-009}\def\cmsCernDate{\today}\def\cmsMessage{Submitted to Physics Letters B}

\begin{document}\cmsNoteHeader{EXO-10-018}

\hyphenation{had-ron-i-za-tion}
\hyphenation{cal-or-i-me-ter}
\hyphenation{de-vices}
\RCS$Revision: 40833 $
\RCS$HeadURL: svn+ssh://alverson@svn.cern.ch/reps/tdr2/papers/EXO-10-018/trunk/EXO-10-018.tex $
\RCS$Id: EXO-10-018.tex 40833 2011-02-21 17:30:52Z alverson $
\cmsNoteHeader{EXO-10-018} 
\title{Search for a Heavy Bottom-like Quark in pp Collisions at $\sqrt{s} = 7$ TeV}

\date{\today}

\abstract{
    A search for pair-produced bottom-like quarks in $\rm pp$ collisions at
    	$\sqrt{s} = 7$ TeV is conducted with the CMS experiment at the LHC.
    The decay $\rm b^\prime \to tW$ is considered in this search.
    The $\rm b^\prime \overline{b}{}^\prime \to tW^- \overline{t}W^+$
    	process can be identified by the distinctive signature of
    trileptons and same-sign dileptons.
    With a data sample corresponding to an integrated luminosity of
    34~pb$^{-1}$, no excess above the standard model background predictions
    is observed and a $\rm b^\prime$ quark with a mass between 255 and
    361~GeV/$c^2$ is excluded at the 95\% confidence level.
}

\hypersetup{%
pdfauthor={CMS Collaboration},%
pdftitle={Search for b' to tW in pp collisions at sqrt(s) = 7 TeV},%
pdfsubject={CMS},%
pdfkeywords={CMS, physics, exotica}}

\maketitle 

The standard model with three generations of quarks
describes remarkably well almost all particle physics phenomena observed to date.
Although adding a fourth generation of massive fermions is
an obvious extension of the model, it
became less popular when limits were obtained on
the number of light neutrino
flavours~\cite{Decamp:1989tu,Aarnio:1989tv,Adeva:1989mn,Akrawy:1989pi,Nash:1989rs}.
In addition,
precise measurements of the electroweak parameters
disfavour such a possibility~\cite{lep2005, Nakamura:2010zzi}.
Recently, however, there has been renewed interest in the
fourth generation~\cite{Holdom:2009rf, Soni:2010xh,
Buras:2010pi, Erler:2010sk, Flacco:2010rg}.
Indirect bounds on the Higgs boson mass can be relaxed~\cite{Frampton:1999xi,Kribs:2007nz}, and
an additional generation of quarks may possess enough
intrinsic matter and anti-matter asymmetry to be relevant
for the baryon asymmetry of the Universe~\cite{Hou:2008xd}.

A search for a heavy bottom-like quark ($\rm b^\prime$) is presented in
pp collisions at a centre-of-mass energy of 7 TeV with the Compact Muon Solenoid (CMS)
detector at the large Hadron Collider (LHC).
The decay chain $\rm b^\prime\overline{b}{}^\prime\to tW^-\overline{t}W^+ \to b W^+W^- \overline{b} W^-W^+$ is expected
to be dominant if the mass of the $\rm b^\prime$ quark ($M_{\rm b^\prime}$) is
larger than the sum of top-quark and W-boson masses~\cite{Arhrib:2006pm}.
A $\rm b^\prime$ mass below this threshold
is disfavoured by results from several previous experiments~\cite{Aaltonen:2009nr}.
As each W boson can decay leptonically into $\mathrm{e}\nu$ or $\mu\nu$ in 22\% of the cases, the full decay chain
may lead to distinctive signatures with two same-sign isolated leptons or
three isolated leptons in the final state,
which covers 7.3\% of the total decays and is expected to happen very rarely in the standard model.
A similar search in these decay channels
has been carried out by the CDF experiment~\cite{Aaltonen:2009nr},
setting a lower limit
of 338~GeV/$c^2$ at the 95\%\ confidence level (CL) on the mass of $\rm b^\prime$
quark\rlap.\footnote{A new analysis from CDF, in the lepton plus multijet
channel, sets the $\rm b^\prime$ quark mass limit at 372~GeV/$c^2$~\cite{Aaltonen:2011vr}.}

The central feature of the CMS detector is a large-solid-angle magnetic spectrometer,
with an axial magnetic field of 3.8 T provided by a superconducting solenoid.
Charged particle trajectories
are measured by a silicon pixel detector and strip tracker.
A lead tungstate crystal electromagnetic calorimeter (ECAL), with a lead-silicon preshower detector in the end-caps,
and a brass/scintillator hadron calorimeter (HCAL) are placed outside of the tracker,
which altogether provide high resolution measurements for electrons/photons and hadronic jets.
The hermetic design of the detector allows good
measurement of missing transverse energy.
Muons are detected by the tracker and a gas-ionization detector embedded in the steel magnetic field return yoke.
A detailed description of the CMS detector can be found
in Ref.~\cite{:2008zzk}.

This analysis is based on a data sample corresponding to an integrated
luminosity of 34 pb$^{-1}$ recorded during 2010 run.
A two-level trigger system~\cite{Adam:2005zf}
selects events for further analysis.
The events in this search are selected
by requiring the presence of at least two electrons
or at least one muon in the trigger.
Given the trigger efficiencies measured in data for single-trigger objects,
the trigger efficiency for the final state in the kinematic region of
the off-line analysis has been determined
to be more than 99\% from simulation studies.

Candidate muons are reconstructed with a global fit of
trajectories using hits in the
tracker and the muon system.
Muons are required to have transverse momenta $p_\mathrm{T} > 20$ GeV/$c$ and $|\eta|<2.4$, where $\eta$ is
the pseudorapidity, defined as $\eta = -\ln[\tan\theta/2]$ and $\theta$ is the polar angle relative to the
counterclockwise proton beam direction as
measured from the nominal interaction vertex.
As discussed in Ref.~\cite{MUOPAS},
the muon candidate must be associated with hits in the silicon
strip and the pixel detector,
the segments in the muon chamber,
and have a high-quality global fit to the track trajectory.
The efficiency for these muon selection criteria is 99\% or higher.
In addition, the track
is required to be consistent with originating from the primary interaction vertex.

Reconstruction of electron candidates
starts from clusters of energy deposits in the ECAL, which are
then matched to hits in the silicon tracker.
Electron candidates are required to have $p_\mathrm{T} > 20$ GeV/$c$.
Candidates are required to be reconstructed in the fiducial volume of the barrel ($|\eta| < 1.478$) or
in the end-caps ($1.55 < |\eta| < 2.4$).
The electron candidate
track is required to be consistent with originating from the interaction vertex.
Electrons are identified using variables which include
the ratio between the energy deposited in the HCAL and the ECAL,
the shower width in $\eta$, and
the distance between the calorimeter shower and
the particle trajectory in the tracker,
measured in both $\eta$ and azimuthal angle ($\phi$).
The selection criteria are optimized~\cite{EGMPAS} to
reject the background from hadronic jets while maintaining an
efficiency of 85\% for the electrons from $\rm W$ or $\rm Z$ decays.

Electrons and muons from $\mathrm{W}\to\ell\nu$ ($\ell=\mathrm{e},\mu$) decays
are expected to be isolated from other particles in the detector.
A cone of $\Delta R < 0.3$,
where $\Delta R = \sqrt{(\Delta\eta)^2 + (\Delta\phi)^2}$,
is constructed around the lepton candidate direction.
The scalar sum of the
track transverse momenta and calorimeter
energy deposits inside the cone projected onto the transverse plane
is calculated, excluding contributions from the lepton candidate.
A barrel (end-cap) electron candidate is rejected
if this scalar sum exceeds 9\% (6\%) of the candidate $p_\mathrm{T}$,
while the scalar sum for a muon candidate is not allowed to exceed 20\%.
Electron candidates are further required to be separated from the selected
muon candidates; any electron candidate within
a $\Delta R < 0.1$ cone of
a muon candidate is rejected to remove misidentified electrons due to muon
bremsstrahlung.
Electron candidates which are identified as coming
from photon conversions are also rejected.

Hadronic jets are clustered from the particles reconstructed with an optimal
use all CMS sub-detectors by the particle-flow global event reconstruction
described in Ref.~\cite{PFT-09-001, PFT-10-001, PFJETPAS, PFT-10-003}, with the anti-$k_\mathrm{T}$ algorithm~\cite{Cacciari:2008gp}.
The energy calibration~\cite{JES} is performed separately for each particle type, the resulting jet energies require only a small correction accounting for thresholds and residual inefficiencies.
Jet candidates are required to have a minimum
$p_\mathrm{T}$ of 25 GeV/$c$ and $|\eta|<2.4$.
Neutrinos from $\rm W$ boson decays
escape the detector and thus produce a significant
energy imbalance in the detector.
An important quantity is the missing transverse
energy, $E\!\!\!/_\mathrm{T}$, which describes
the imbalance of detected energy
perpendicular to the beam direction.
It is determined as the vectorial sum of the transverse momenta of all
particles reconstructed by the particle-flow algorithm~\cite{MET, PFJETPAS}.

Events are required
to have at least one well reconstructed interaction
vertex~\cite{2010EPJC..tmp..299K}.
Events with two same-sign leptons
or with three leptons (with two of them are oppositely charged)
are selected.
Events with fewer than four (two) jets are rejected
for the same-sign dilepton (trilepton) channel.
In addition, events with an oppositely-charged muon or electron pair
with $|M_{\ell^+\ell^-}-M_Z| < 10$ GeV/$c^2$ are rejected in order to suppress
the background from $Z$ decays. The backgrounds due to charge misidentification
are substantially larger for electrons than muons, thus events with same-sign
electron pair with $|M_{\rm e^\pm e^\pm}-M_\mathrm{Z}| < 10$ GeV/$c^2$  are also discarded.
For each event, the scalar quantity
$S_\mathrm{T} = \sum p_\mathrm{T} ({\rm jets}) + \sum p_\mathrm{T} ({\rm leptons}) + {E\!\!\!/_\mathrm{T}}$
is determined and a minimum $S_\mathrm{T}$ of
350 GeV is required.

\begin{figure}[tp]
  \begin{center}
    \includegraphics[width=0.45\columnwidth]{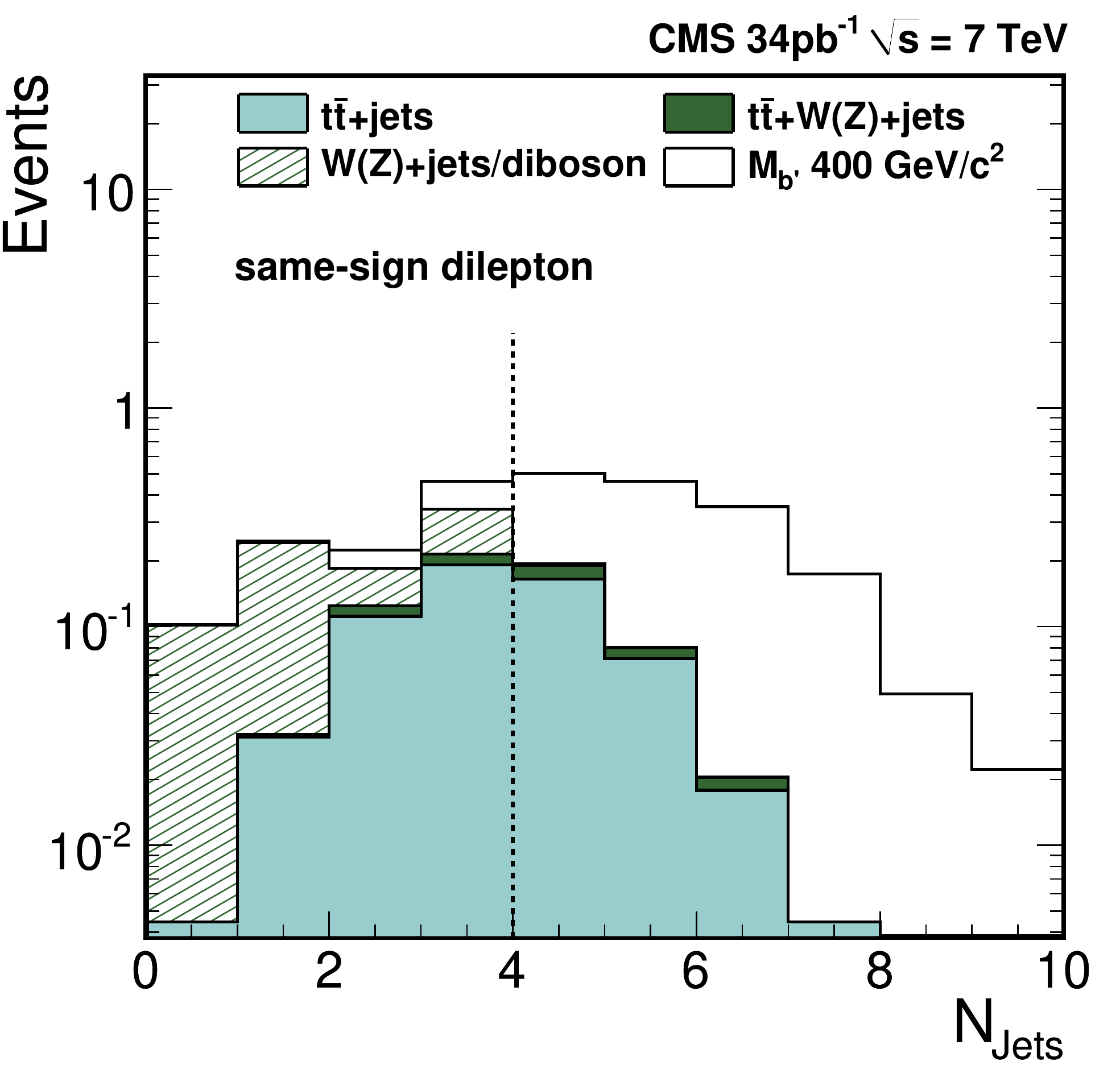}
    \includegraphics[width=0.45\columnwidth]{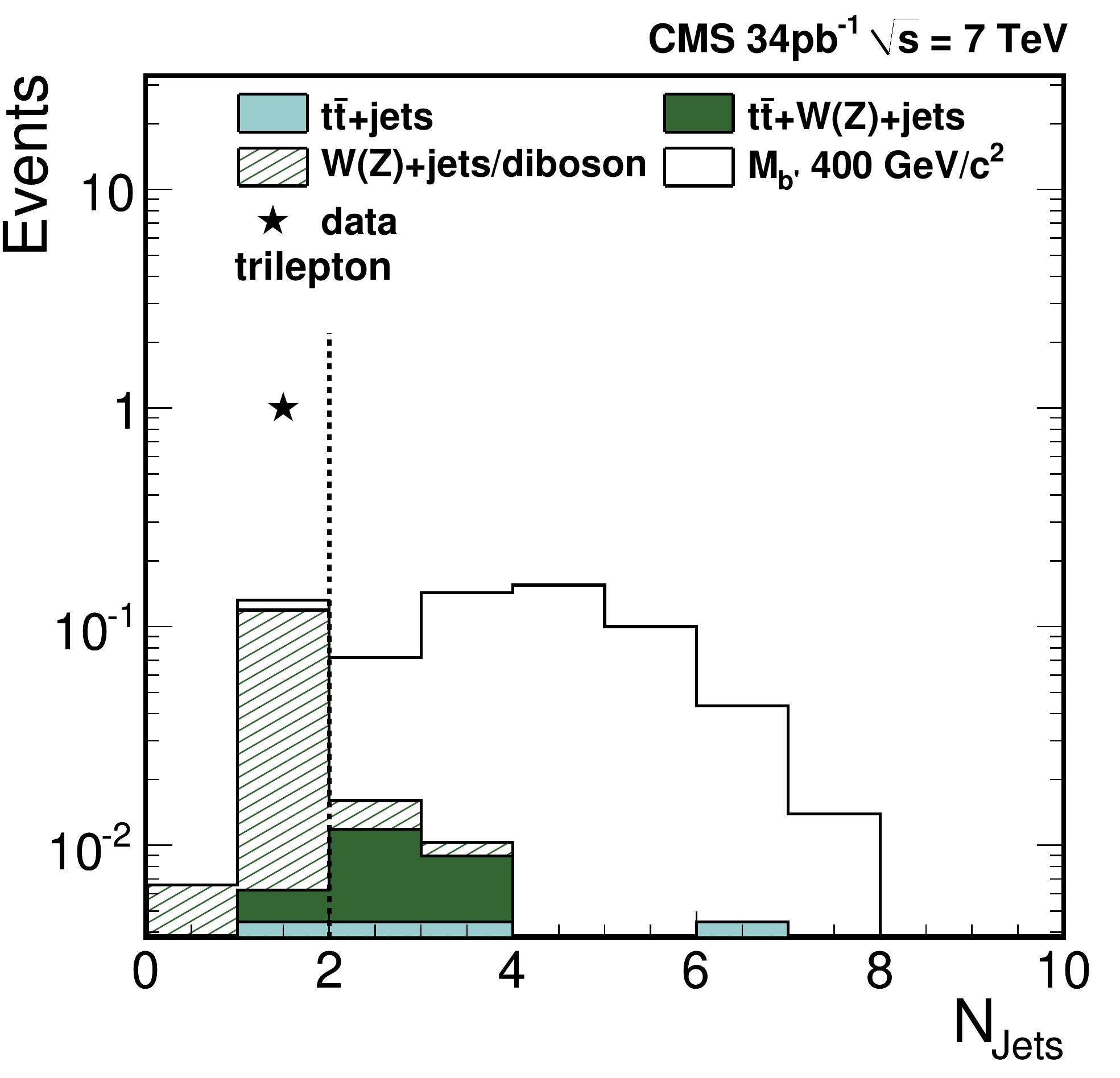}
    \caption{
Jet multiplicity distributions for the same-sign dilepton channel (left), and the trilepton channel (right). The star in the right plot represents the single measured event, which fails to satisfy the requirement on jet multiplicity. The open histogram is the signal contribution expected from a $\rm b^\prime$ with $M_{\rm b^\prime} =$ 400 GeV/$c^2$. The light blue and dark green filled histograms show the contributions from $\rm t\overline{t} + jets$ and $\rm t\overline{t}+W(Z)+jets$ respectively. The shaded histogram represents electroweak processes ($\rm W(Z)+jets$, dibosons). All selections are applied except the one corresponding to the plotted variable. The vertical dotted lines indicate the minimum numbers of jets required in events selected for each of the channels.
    }
    \label{fig:jet_multiplicities}
  \end{center}
\end{figure}

\begin{figure}[tp]
  \begin{center}
    \includegraphics[width=0.45\columnwidth]{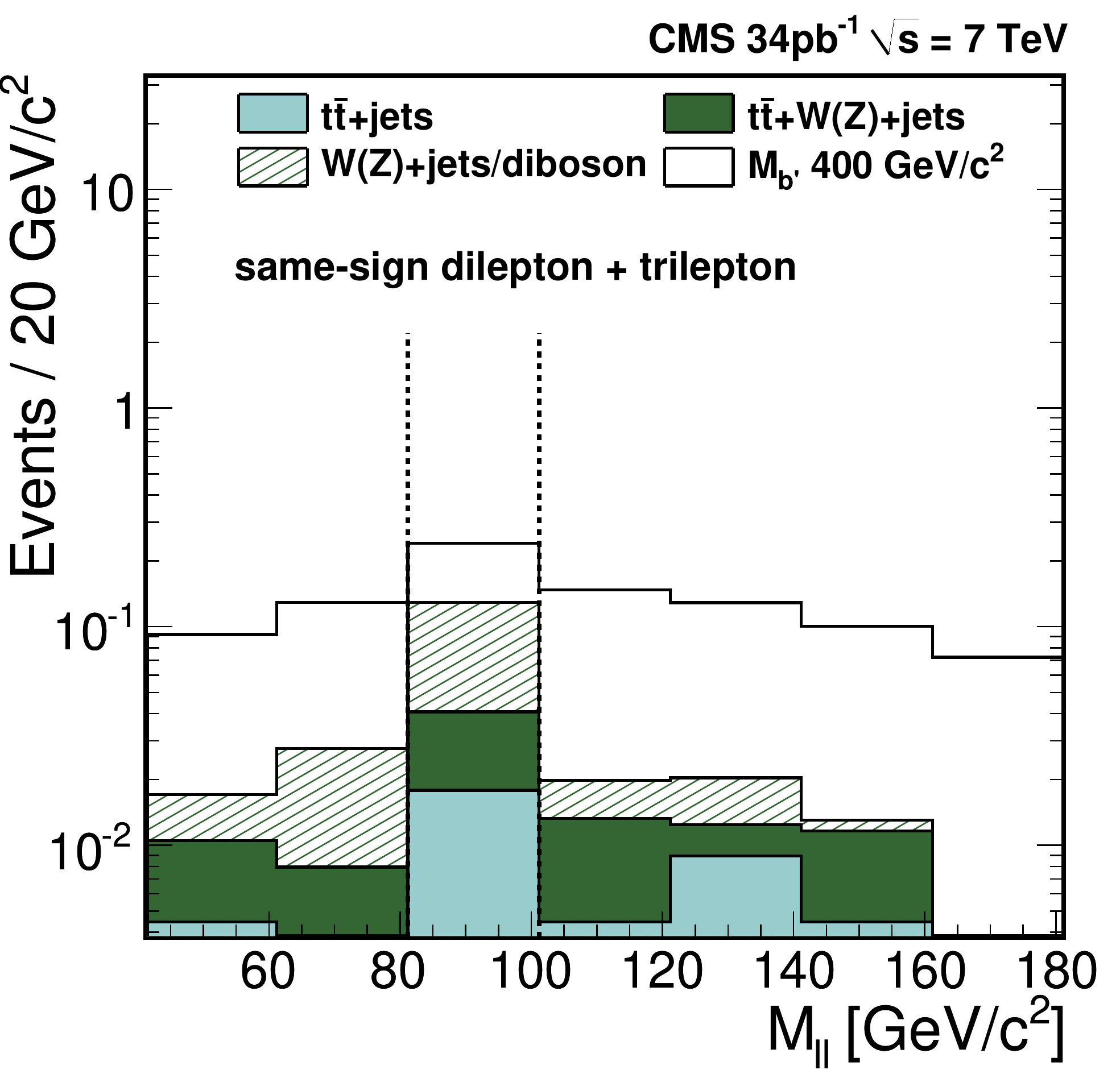}
    \includegraphics[width=0.45\columnwidth]{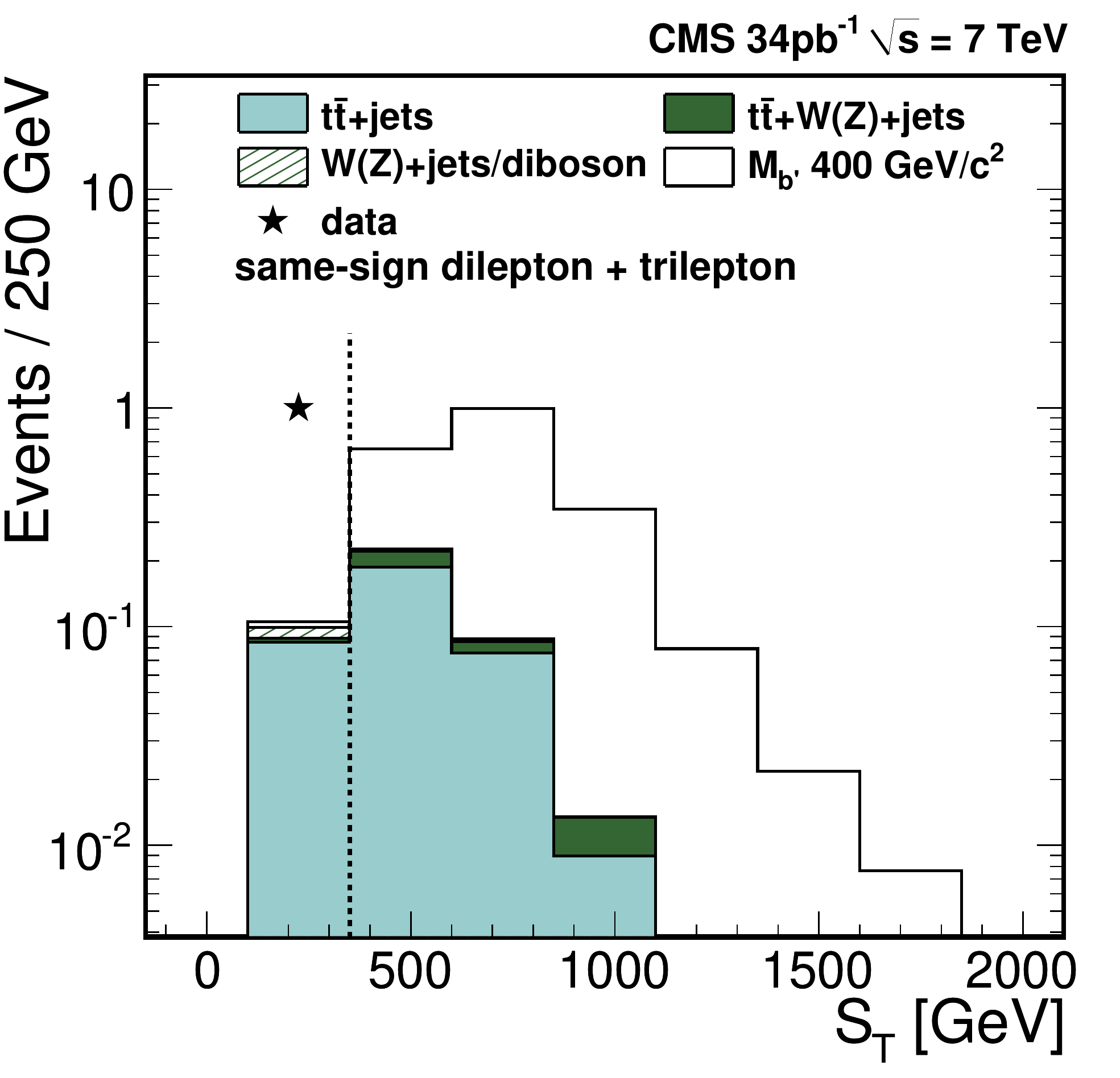}
    \caption{The invariant mass distribution (left) of two muons with opposite charges or
    electrons of any charge, $M(\ell\ell)$, and
    the $S_\mathrm{T}$ distribution (right) including same-sign dilepton and trilepton channels.
    The star in the right plot represents the measured event, which fails to satisfy the requirement on $S_\mathrm{T}$.
    The open histogram is the signal contribution expected from a $\rm b^\prime$ with $M_{\rm b^\prime} =$ 400 GeV/$c^2$.
    The light blue and dark green filled histograms show the contributions from $\rm t\overline{t} + jets$
    and $\rm t\overline{t}+W(Z)+jets$ respectively.
    The shaded histogram represents electroweak processes ($\rm W(Z)+jets$, dibosons).
	All selections are applied except the one corresponding to the plotted variable.
    Events with an electron pair or an opposite sign muon pair, with $M(\ell\ell)$ falling in the
    region defined by the vertical dotted lines on the left plot, are rejected in order to suppress
    the background from Z events.
    The vertical dotted line in the right plot indicates the lower $S_\mathrm{T}$ threshold used in the analysis.
    }
    \label{fig:misc_plots}
  \end{center}
\end{figure}

Selection efficiencies for signal events are estimated using samples
simulated with the
{\sc MadGraph/MadEvent} generator (v4.4.26)~\cite{Maltoni:2002qb} with up to
two additional partons in the hard interactions.
The events are subsequently processed with {\sc pythia} (v6.420)~\cite{PYTHIA}
to provide
parton showering and hadronization of the particles,
and then passed through a simulation of the
CMS detector based on {\sc geant4}~\cite{Agostinelli:2002hh}.
The signal efficiency varies from 3.1 to 4.6\% for $\rm b^\prime$
masses between 300 and 500 GeV/$c^2$.
These efficiencies include the
W decay branching fractions.
The jet multiplicities for the trilepton and same-sign
dilepton channels are shown
in Fig.~\ref{fig:jet_multiplicities}.
The distributions of dilepton invariant mass $M_{\ell\ell}$ and $S_\mathrm{T}$ are
presented in Fig.~\ref{fig:misc_plots}.
The expected distributions of the $\rm b^\prime$ signal are normalized with
the production cross section calculated at the next-to-leading order (NLO)
in $\alpha_s$~\cite{Berger:2009qy}, for a $\rm b^\prime$ with 400 GeV/$c^2$ mass.

The expected yields and efficiencies for signal and background
are summarized in Table~\ref{tab:yield_expected}.
The background contributions
from $\rm pp \to t\overline{t} + jets$ and $\rm W/Z + jets$
are normalized to the CMS measured
inclusive $\rm pp \to t\overline{t}$, $\rm W$, and $Z$
cross sections~\cite{Khachatryan:2010ez, Khachatryan:2010xn}.
The simulated samples for $\rm pp \to t\overline{t} + jets$ and $\rm W/Z + jets$ processes include
initial state $\rm b$ and $\rm c$ quarks in the hard interactions.
Production of dibosons is
estimated with NLO cross sections given by {\sc mcfm}~\cite{Campbell:2010ff}.
The $\rm t\overline{t}+W/Z$ and
same-sign $\rm WW+jj$ processes are calculated
using the {\sc MadGraph} generator at leading order (LO) in $\alpha_s$.
The total background yield is estimated to be 0.33.
The only dominant background contribution comes from
$\rm pp \to t\overline{t} + jets$ events;
contributions from other processes are very small.

  \begin{table*}[t]
    \caption{Summary of expected signal and background production cross sections,
    selection efficiencies $\epsilon$, expected yields,
    and the observed event yield in data.
    The cross sections are obtained from leading order predictions, next-to-leading order predictions,
    or CMS measurements.}
    \label{tab:yield_expected}
    \begin{center}
      \begin{tabular}{|l|c|c|c|} \hline
        Process & Cross section & $\epsilon$ [\%] & Yield \\
        \hline
        $\rm b^\prime\overline{b}{}^\prime$, $M_{\mathrm{b^\prime}} = 300$ GeV/$c^2$ & 7.29 pb (NLO)  & 3.08 & 7.7  \\
        $\rm b^\prime\overline{b}{}^\prime$, $M_{\mathrm{b^\prime}} = 350$ GeV/$c^2$ & 2.94 pb (NLO)  & 3.75 & 3.8  \\
        $\rm b^\prime\overline{b}{}^\prime$, $M_{\mathrm{b^\prime}} = 400$ GeV/$c^2$ & 1.30 pb (NLO)  & 3.99 & 1.8  \\
        $\rm b^\prime\overline{b}{}^\prime$, $M_{\mathrm{b^\prime}} = 450$ GeV/$c^2$ & 0.617 pb (NLO) & 4.34 & 0.91 \\
        $\rm b^\prime\overline{b}{}^\prime$, $M_{\mathrm{b^\prime}} = 500$ GeV/$c^2$ & 0.310 pb (NLO) & 4.58 & 0.49 \\
        \hline
        $\rm t\overline{t}+jets$     & $1.9\times10^2$ pb (CMS)     & $4.1\times10^{-3}$ &  0.27   \\ 
        $\rm t\overline{t}+W+jets$           & 0.144 pb (LO)  & $0.67$ &  0.033    \\
        $\rm t\overline{t}+Z+jets$           & 0.094 pb (LO)  & $0.50$ &  0.016   \\
        $\rm W+jets$                & $3.0\times10^4$ pb (CMS)        & $<1.0\times10^{-5}$ &  $<0.11$   \\ 
        $\rm Z+jets$                & $2.9\times10^3$ pb (CMS)        & $<9.2\times10^{-5}$ &  $<0.09$   \\ 
        $\rm WW$                       & 43 pb (NLO)     & $<8.2\times10^{-4}$ &  $<0.012$   \\
        $\rm WZ$                       & 18 pb (NLO)     & $<8.1\times10^{-4}$ &  $<0.005$   \\
        $\rm ZZ$                       & 5.9 pb (NLO)    & $3.0\times10^{-3}$  &  $0.006 $   \\
        Same-sign $\rm WW+jj$          & 0.15 pb (LO)    & $3.9\times10^{-2}$  &  $0.002 $   \\
        \hline
        Background sum              & - & - & $0.33 $ \\
        \hline
    Data-driven background yield    & - & - & $0.32 $ \\
        \hline
    Observed yield in data          & -  & - & 0 \\
        \hline
      \end{tabular}
    \end{center}
  \end{table*}

For the same-sign dilepton channel,
there are two types of $\rm t\overline{t}$
background:
single-lepton $\rm t\overline{t}$ events
with an extra misidentified or non-isolated lepton,
or dilepton $\rm t\overline{t}$
events with a charge-misidentified electron.
Backgrounds are estimated from data as follows.

Leptons chosen with
relaxed selection criteria
are denoted as ``loose'' muon or ``loose'' electron.
Leptons chosen with the full selection criteria defined above are
denoted as ``tight'' muons and ``tight'' electrons.
The background events with a misidentified or non-isolated
lepton are estimated using a control region with
one tight lepton and one loose lepton,
with the rest of the selection criteria exactly the same as for signal.
The background contribution is calculated
from the yields observed in the control regions
multiplied by the ratios of the number of electrons
or muons passing
tight and loose cuts.
These ratios
are determined from data by taking the
ratios between the number of events in the control
region with two loose leptons,
and the control region
with one loose plus one tight lepton.
The background contribution from
electron
charge misidentification is determined
from control regions with
oppositely-charged electron pairs or from e-$\mu$ events.
The charge misidentification rate ($0.6\pm0.1$\%)
is determined by measuring the $\rm Z$ boson events
reconstructed using two electron candidates with the same electric charge, and
is normalized to the yield of $\rm Z \to e^+e^-$ events.

For the trilepton channel,
the background yield in the signal region
is estimated using a control
region with the same criteria as for
the signal, but requiring only two leptons with opposite charges.
The normalization between the background
in the signal region and the background in the
control region is determined from simulations.

The background yield in the signal region, including both trilepton and same-sign dilepton channels,
is estimated to be 0.32.
The systematic uncertainties on the $\rm t\overline{t}$ background estimations for the same-sign dilepton channel were evaluated
using a mixture of simulated samples.
The normalization for each physics process in the simulated events
is derived from the cross sections.
Applying this estimation procedure to the samples of simulated events gives an estimated background of 0.21 events.
This is in good agreement with the figure of 0.33 events obtained by counting directly the number of simulated background events
satisfying the signal selection.
The difference between these two yields is included in the systematic uncertainties.
The background estimation procedure for trilepton events is assumed to have a systematic uncertainty of $\pm$100\%
on the simulated normalization ratio.
The sum of these two uncertainties, which
arise from the bias of control-region methods,
provide the dominant uncertainty of 56\% on the background yield.

The relative uncertainty on the
integrated luminosity measurement is estimated to be 11\%~\cite{lumi} and
is included in the limit calculations.
The effect of this uncertainty in the background estimation cancels when
the absolute normalization of backgrounds
are taken from the measured yields in the control regions.
The statistical uncertainties
on the yields in the control regions are included in the uncertainty
on the backgrounds.
The QCD multijet contribution is estimated to be smaller then 0.09 events,
and considered as a systematic uncertainty of 29\% on background estimation.
The uncertainties on the background cross sections
are included by varying the
normalization on the relevant processes
as follows:
$\pm39$\% for $\rm t\overline{t}+ jets$~\cite{Khachatryan:2010ez},
$\pm3$\% ($\pm4$\%) for $\rm W$ ($\rm Z$)~\cite{Khachatryan:2010xn},
$\pm$(27 to 42)\% for dibosons, and $\pm50$\% for other processes.
Lepton selection efficiencies are measured using
inclusive $\rm Z$ samples; the resulting differences between data and simulated samples are
smaller than 2\%.
An additional systematic uncertainty was assigned with a magnitude of 50\% on the efficiency
difference between simulated $\rm Z$ and $\rm b^\prime$ samples due to
the effects of different event topologies.
This results in 5.8\% and 5.4\% uncertainties
for the electrons and muons, respectively.
Weighted averages
including trilepton and same-sign dilepton final states in the appropriate proportions of selected muons and electrons
result uncertainties of 13\% and 1.5\% in signal efficiency and background estimation, respectively.
Uncertainty sets given by CTEQ6~\cite{Pumplin:2002vw} are used to determine the uncertainties from parton distribution functions (PDFs). Weights for each simulated event are recalculated,
and the variations are summed in quadrature.
The systematic effects of the jet energy scale
uncertainty, jet resolution, $E\!\!\!/_\mathrm{T}$ resolution,
and jets from pile-up are found to be small~\cite{JES, MET}.
The total uncertainties on the signal selection efficiency and on the
background estimation are evaluated to be 13\% and 65\%, respectively,
and are summarized in Table~\ref{tab:systematic_error}.

  \begin{table*}[tp]
    \caption{Summary of relative systematic uncertainties for signal selection efficiencies ($\Delta\epsilon/\epsilon$) and for background estimations ($\Delta B/B$). The ranges represent the dependence on the input $\rm b^\prime$ mass.}
    \label{tab:systematic_error}
    \begin{center}
      \begin{tabular}{|l|c|c|} \hline
                                  & $\Delta\epsilon/\epsilon$ [\%] & $\Delta B/B$ [\%]  \\	
	\hline
	Accuracy of control-region method & - & 56   \\
	\hline
	Norm: QCD multijet           & - & 29   \\
	Norm: $\rm t\overline{t}+ jets$  & - & 0.5  \\
	Norm: $\rm W(Z)+ jets$       & - & 1.0  \\
	Norm: dibosons               & - & 0.9  \\
	Norm: other processes        & - & 5.5  \\
	\hline							       	
	Jet energy scale             & 1.1 -- 2.1 & 1.0  \\
	Jet energy resolution        & 0.1 -- 0.6 & 1.5  \\
	Missing energy resolution    & 0.1 -- 1.2 & 5.6  \\
	Lepton selection             & 13         & 1.5  \\	
	Pile-up                      & 1.0 -- 1.2 & $<0.1$  \\
	PDF                          & 0.5 -- 1.0 & 1.0  \\
	Control region statistics    & -          & 13   \\
	Simulated sample statistics  & 2.4 -- 3.0 & -      \\
	\hline
	Total                        & 13         & 65   \\
	\hline
      \end{tabular}
    \end{center}

  \end{table*}

The background yield in the signal region is
0.32 with a total relative uncertainty of 65\%.
No events are observed in the data, which is consistent
with the background expectation.
An event is found below the $S_\mathrm{T}$ threshold in the same-sign dilepton channel (Fig.~\ref{fig:misc_plots}), and
another event is rejected by the jet multiplicity requirement in the trilepton channel (Fig.~\ref{fig:jet_multiplicities}).
These two events are consistent with the expected total background yield of 0.69,
if the requirements on $S_\mathrm{T}$ and jet multiplicity in trilepton channel are
relaxed to $200$~GeV/$c^2$ and $1$, respectively.

\begin{figure}[tp]
  \begin{center}
    \includegraphics[width=0.95\textwidth]{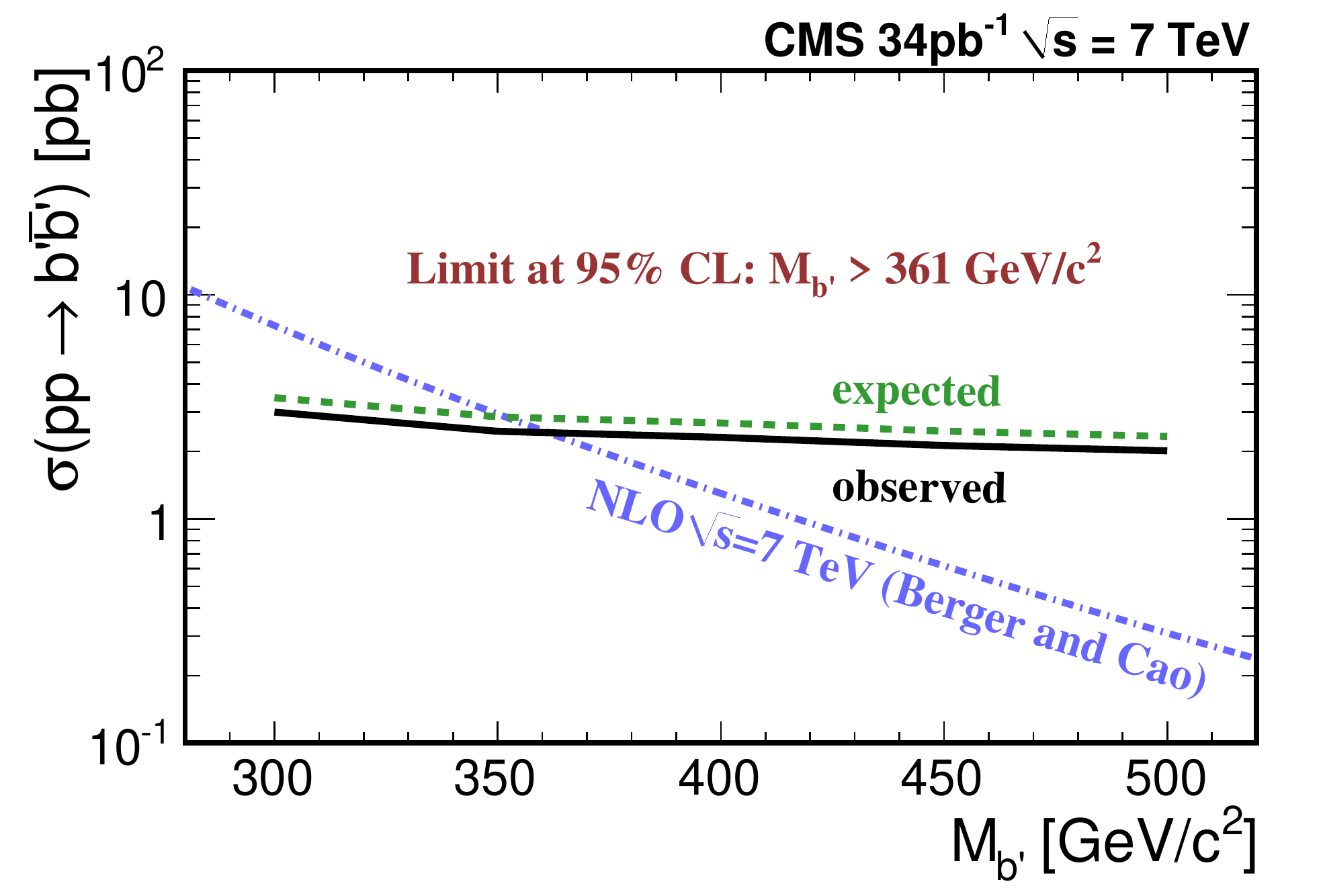}
    \caption{The exclusion limits at the 95\% CL on the $\rm pp \to b^\prime\overline{b}{}^\prime$
    production cross section.
    The solid line represents the observed limits,
    while the dotted line represents
    the limit expected with the available sample size, assuming the presence of standard model processes alone.
    Comparing with NLO production cross sections, $\rm b^\prime$ mass less than 361 GeV/$c^2$ is excluded.
    }
    \label{fig:s95exclusion}
  \end{center}
\end{figure}

For each $\rm b^\prime$ mass hypothesis, cross sections, selection efficiencies and associated uncertainties are estimated
(table~\ref{tab:yield_expected}). From these, from the estimated background yield and zero selected events, upper limits
on $\rm b^\prime\overline{b}{}^\prime$ cross sections
at the 95\% CL are derived using a Bayesian method
with a log-normal prior for integration over the nuisance parameters~\cite{Bayes}.
The resulting upper limits obtained on the $\rm b^\prime$ cross section are 3.00, 2.46, 2.31, 2.13, and 2.01 pb for
mass hypotheses of 300, 350, 400, 450, and 500 GeV/$c^2$, respectively.
These limits are plotted as
the solid line in Fig.~\ref{fig:s95exclusion}, while the dotted line
represents the limit expected with the available sample size, assuming the presence of standard model processes alone.
By comparing to the NLO production cross section
for $\rm pp \to b^\prime\overline{b}{}^\prime$,
a lower limit of 361 GeV/$c^2$ is extracted for the mass of the $\rm b^\prime$ quark at the 95\% CL.

In summary, a search for a heavy bottom-like quark
produced in proton-proton collisions at $\sqrt{s} = 7$ TeV has been presented.
The production of $\rm pp \to b^\prime \overline{b}{}^\prime \to t\overline{t}W^+W^-$
has been studied in a data set corresponding to an integrated luminosity of 34 pb$^{-1}$
collected by the CMS detector during 2010.
Final states with the signatures of trileptons or same-sign dileptons
are very rare in standard model processes, and
background contributions have been estimated to be very small.
No events are found in the signal region defined in the analysis, and
the $\rm b^\prime$ mass range from 255 to 361 GeV/$c^2$ has been excluded at the 95\% CL.

We wish to congratulate our colleagues in the CERN accelerator departments for the excellent performance of the LHC machine. We thank the technical and administrative staff at CERN and other CMS institutes, and acknowledge support from: FMSR (Austria); FNRS and FWO (Belgium); CNPq, CAPES, FAPERJ, and FAPESP (Brazil); MES (Bulgaria); CERN; CAS, MoST, and NSFC (China); COLCIENCIAS (Colombia); MSES (Croatia); RPF (Cyprus); Academy of Sciences and NICPB (Estonia); Academy of Finland, ME, and HIP (Finland); CEA and CNRS/IN2P3 (France); BMBF, DFG, and HGF (Germany); GSRT (Greece); OTKA and NKTH (Hungary); DAE and DST (India); IPM (Iran); SFI (Ireland); INFN (Italy); NRF and WCU (Korea); LAS (Lithuania); CINVESTAV, CONACYT, SEP, and UASLP-FAI (Mexico); PAEC (Pakistan); SCSR (Poland); FCT (Portugal); JINR (Armenia, Belarus, Georgia, Ukraine, Uzbekistan); MST and MAE (Russia); MSTD (Serbia); MICINN and CPAN (Spain); Swiss Funding Agencies (Switzerland); NSC (Taipei); TUBITAK and TAEK (Turkey); STFC (United Kingdom); DOE and NSF (USA).

\bibliography{auto_generated}   

\providecommand{\href}[2]{#2}\begingroup\raggedright\begin{thebibliography}{10}%
\makeatletter
\providecommand{\hrefCMSnoop }[0]{\@secondoftwo}%
\makeatother

\bibitem{Decamp:1989tu}
\hrefCMSnoop {} {{ ALEPH} Collaboration, ``Determination of the Number of Light
  Neutrino Species'',} \textit{ Phys. Lett.} \textbf{ B231} (1989) 519.
\href{http://dx.doi.org/10.1016/0370-2693(89)90704-1}{\texttt{
  doi:10.1016/0370-2693(89)90704-1}}.

\bibitem{Aarnio:1989tv}
\hrefCMSnoop {} {{ DELPHI} Collaboration, ``Measurement of the Mass and Width
  of the $\rm {Z}^0$ Particle from Multi-Hadronic Final States Produced in $\rm
  e^{+} e^{-}$ Annihilations'',} \textit{ Phys. Lett.} \textbf{ B231} (1989)
  539.
\href{http://dx.doi.org/10.1016/0370-2693(89)90706-5}{\texttt{
  doi:10.1016/0370-2693(89)90706-5}}.

\bibitem{Adeva:1989mn}
\hrefCMSnoop {} {{ L3} Collaboration, ``A Determination of the Properties of
  the Neutral Intermediate Vector Boson $\rm {Z}^0$'',} \textit{ Phys. Lett.}
  \textbf{ B231} (1989) 509.
\href{http://dx.doi.org/10.1016/0370-2693(89)90703-X}{\texttt{
  doi:10.1016/0370-2693(89)90703-X}}.

\bibitem{Akrawy:1989pi}
\hrefCMSnoop {} {{ OPAL} Collaboration, ``Measurement of the $\rm {Z}^0$ Mass
  and Width with the {OPAL} Detector at {LEP}'',} \textit{ Phys. Lett.}
  \textbf{ B231} (1989) 530.
\href{http://dx.doi.org/10.1016/0370-2693(89)90705-3}{\texttt{
  doi:10.1016/0370-2693(89)90705-3}}.

\bibitem{Nash:1989rs}
\hrefCMSnoop {} {{ MARK II} Collaboration, ``A Measurement of The {Z} Boson
  Resonance Parameters at the {SLC}'',} \textit{ International Europhysics
  Conference on High Energy Physics} \textbf{ SLAC-PUB-5141} (1989).

\bibitem{lep2005}
\hrefCMSnoop {} {{ ALEPH, DELPHI, L3, OPAL and SLD} Collaboration, ``Precision
  electroweak measurements on the {Z} resonance'',} \textit{ Phys. Rept.}
  \textbf{ 427} (2006) 257,
  \href{http://www.arXiv.org/abs/hep-ex/0509008}{\texttt{
  arXiv:hep-ex/0509008}}.
\href{http://dx.doi.org/10.1016/j.physrep.2005.12.006}{\texttt{
  doi:10.1016/j.physrep.2005.12.006}}.

\bibitem{Nakamura:2010zzi}
\hrefCMSnoop {} {{ Particle Data Group} Collaboration, ``Review of particle
  physics'',} \textit{ J. Phys.} \textbf{ G37} (2010) 075021.
\href{http://dx.doi.org/10.1088/0954-3899/37/7A/075021}{\texttt{
  doi:10.1088/0954-3899/37/7A/075021}}.

\bibitem{Holdom:2009rf}
\hrefCMSnoop {} {B.~Holdom {et~al.}, ``Four Statements about the Fourth
  Generation'',} \textit{ PMC Phys.} \textbf{ A3} (2009) 4,
  \href{http://www.arXiv.org/abs/0904.4698}{\texttt{ arXiv:0904.4698}}.
\href{http://dx.doi.org/10.1186/1754-0410-3-4}{\texttt{
  doi:10.1186/1754-0410-3-4}}.

\bibitem{Soni:2010xh}
\hrefCMSnoop {} {A.~Soni {et~al.}, ``{SM} with four generations: Selected
  implications for rare {B} and {K} decays'',} \textit{ Phys. Rev.} \textbf{
  D82} (2010) 033009, \href{http://www.arXiv.org/abs/1002.0595}{\texttt{
  arXiv:1002.0595}}.
\href{http://dx.doi.org/10.1103/PhysRevD.82.033009}{\texttt{
  doi:10.1103/PhysRevD.82.033009}}.

\bibitem{Buras:2010pi}
\hrefCMSnoop {} {A.~J. Buras {et~al.}, ``Patterns of Flavour Violation in the
  Presence of a Fourth Generation of Quarks and Leptons'',} \textit{ JHEP}
  \textbf{ 09} (2010) 106, \href{http://www.arXiv.org/abs/1002.2126}{\texttt{
  arXiv:1002.2126}}.
\href{http://dx.doi.org/10.1007/JHEP09(2010)106}{\texttt{
  doi:10.1007/JHEP09(2010)106}}.

\bibitem{Erler:2010sk}
\hrefCMSnoop {} {J.~Erler and P.~Langacker, ``Precision Constraints on Extra
  Fermion Generations'',} \textit{ Phys. Rev. Lett.} \textbf{ 105} (2010)
  031801, \href{http://www.arXiv.org/abs/1003.3211}{\texttt{ arXiv:1003.3211}}.
\href{http://dx.doi.org/10.1103/PhysRevLett.105.031801}{\texttt{
  doi:10.1103/PhysRevLett.105.031801}}.

\bibitem{Flacco:2010rg}
\hrefCMSnoop {} {C.~J. Flacco {et~al.}, ``Direct Mass Limits for Chiral
  Fourth-Generation Quarks in All Mixing Scenarios'',} \textit{ Phys. Rev.
  Lett.} \textbf{ 105} (2010) 111801,
  \href{http://www.arXiv.org/abs/1005.1077}{\texttt{ arXiv:1005.1077}}.
\href{http://dx.doi.org/10.1103/PhysRevLett.105.111801}{\texttt{
  doi:10.1103/PhysRevLett.105.111801}}.

\bibitem{Frampton:1999xi}
\hrefCMSnoop {} {P.~H. Frampton, P.~Q. Hung, and M.~Sher, ``Quarks and leptons
  beyond the third generation'',} \textit{ Phys. Rept.} \textbf{ 330} (2000)
  263, \href{http://www.arXiv.org/abs/hep-ph/9903387}{\texttt{
  arXiv:hep-ph/9903387}}.
\href{http://dx.doi.org/10.1016/S0370-1573(99)00095-2}{\texttt{
  doi:10.1016/S0370-1573(99)00095-2}}.

\bibitem{Kribs:2007nz}
\hrefCMSnoop {} {G.~D. Kribs {et~al.}, ``Four generations and {Higgs}
  physics'',} \textit{ Phys. Rev.} \textbf{ D76} (2007) 075016,
  \href{http://www.arXiv.org/abs/0706.3718}{\texttt{ arXiv:0706.3718}}.
\href{http://dx.doi.org/10.1103/PhysRevD.76.075016}{\texttt{
  doi:10.1103/PhysRevD.76.075016}}.

\bibitem{Hou:2008xd}
\hrefCMSnoop {} {W.-S. Hou, ``{CP} Violation and Baryogenesis from New Heavy
  Quarks'',} \textit{ Chin. J. Phys.} \textbf{ 47} (2009) 134,
\href{http://www.arXiv.org/abs/0803.1234}{\texttt{ arXiv:0803.1234}}.

\bibitem{Arhrib:2006pm}
\hrefCMSnoop {} {A.~Arhrib and W.-S. Hou, ``Flavor changing neutral currents
  involving heavy quarks with four generations'',} \textit{ JHEP} \textbf{ 07}
  (2006) 009, \href{http://www.arXiv.org/abs/hep-ph/0602035}{\texttt{
  arXiv:hep-ph/0602035}}.
\href{http://dx.doi.org/10.1088/1126-6708/2006/07/009}{\texttt{
  doi:10.1088/1126-6708/2006/07/009}}.

\bibitem{Aaltonen:2009nr}
\hrefCMSnoop {} {{ CDF} Collaboration, ``Search for New Bottomlike Quark Pair
  Decays $Q\overline{Q} \to (tW^{\mp})(\overline{t}W^{\pm})$ in Same-Charge
  Dilepton Events'',} \textit{ Phys. Rev. Lett.} \textbf{ 104} (2010) 091801,
  \href{http://www.arXiv.org/abs/0912.1057}{\texttt{ arXiv:0912.1057}}.
\href{http://dx.doi.org/10.1103/PhysRevLett.104.091801}{\texttt{
  doi:10.1103/PhysRevLett.104.091801}}.

\bibitem{Aaltonen:2011vr}
\hrefCMSnoop {} {{ CDF} Collaboration, ``Search for heavy bottom-like quarks
  decaying to an electron or muon and jets in $p\bar{p}$ collisions at
  $\sqrt{s}=1.96$ {TeV}'',}
\href{http://www.arXiv.org/abs/1101.5728}{\texttt{ arXiv:1101.5728}}.

\bibitem{:2008zzk}
\hrefCMSnoop {} {{ CMS} Collaboration, ``The {CMS} experiment at the {CERN
  LHC}'',} \textit{ JINST} \textbf{ 3} (2008) S08004.
\href{http://dx.doi.org/10.1088/1748-0221/3/08/S08004}{\texttt{
  doi:10.1088/1748-0221/3/08/S08004}}.

\bibitem{Adam:2005zf}
\hrefCMSnoop {} {{ CMS Trigger and Data Acquisition Group} Collaboration, ``The
  {CMS} high level trigger'',} \textit{ Eur. Phys. J.} \textbf{ C46} (2006)
  605--667, \href{http://www.arXiv.org/abs/hep-ex/0512077}{\texttt{
  arXiv:hep-ex/0512077}}.
\href{http://dx.doi.org/10.1140/epjc/s2006-02495-8}{\texttt{
  doi:10.1140/epjc/s2006-02495-8}}.

\bibitem{MUOPAS}
\href {http://cdsweb.cern.ch/record/1279140} {{ CMS} Collaboration,
  ``Performance of muon identification in pp collisions at $\sqrt{s}$ = 7
  {TeV}'',} \textit{ CMS Physics Analysis Summary} \textbf{
  \href{http://cdsweb.cern.ch/record/1279140}{CMS-PAS-MUO-10-002}} (2010).

\bibitem{EGMPAS}
\href {http://cdsweb.cern.ch/record/1299116} {{ CMS} Collaboration, ``Electron
  reconstruction and identification at $\sqrt{s}$=7 {TeV}'',} \textit{ CMS
  Physics Analysis Summary} \textbf{
  \href{http://cdsweb.cern.ch/record/1299116}{CMS-PAS-EGM-10-004}} (2010).

\bibitem{PFT-09-001}
\href {http://cdsweb.cern.ch/record/1194487} {{ CMS} Collaboration,
  ``Particle-Flow Event Reconstruction in {CMS} and Performance for Jets, Taus,
  and {MET}'',} \textit{ CMS Physics Analysis Summary} \textbf{
  \href{http://cdsweb.cern.ch/record/1194487}{CMS-PAS-PFT-09-001}} (2009).

\bibitem{PFT-10-001}
\href {http://cdsweb.cern.ch/record/1247373} {{ CMS} Collaboration,
  ``Commissioning of the Particle-flow Event Reconstruction with the first LHC
  collisions recorded in the CMS detector'',} \textit{ CMS Physics Analysis
  Summary} \textbf{
  \href{http://cdsweb.cern.ch/record/1247373}{CMS-PAS-PFT-10-001}} (2010).

\bibitem{PFJETPAS}
\href {http://cdsweb.cern.ch/record/1279341} {{ CMS} Collaboration,
  ``Commissioning of the Particle-Flow reconstruction in Minimum-Bias and Jet
  Events from pp Collisions at 7 TeV'',} \textit{ CMS Physics Analysis Summary}
  \textbf{ \href{http://cdsweb.cern.ch/record/1279341}{CMS-PAS-PFT-10-002}}
  (2010).

\bibitem{PFT-10-003}
\href {http://cdsweb.cern.ch/record/1279347} {{ CMS} Collaboration,
  ``Particle-flow commissioning with muons and electrons from {J/Psi} and {W}
  events at 7 {TeV}'',} \textit{ CMS Physics Analysis Summary} \textbf{
  \href{http://cdsweb.cern.ch/record/1279347}{CMS-PAS-PFT-10-003}} (2010).

\bibitem{Cacciari:2008gp}
\hrefCMSnoop {} {M.~Cacciari, G.~P. Salam, and G.~Soyez, ``The anti-$k_t$ jet
  clustering algorithm'',} \textit{ JHEP} \textbf{ 04} (2008) 063,
  \href{http://www.arXiv.org/abs/0802.1189}{\texttt{ arXiv:0802.1189}}.
\href{http://dx.doi.org/10.1088/1126-6708/2008/04/063}{\texttt{
  doi:10.1088/1126-6708/2008/04/063}}.

\bibitem{JES}
\href {http://cdsweb.cern.ch/record/1308178} {{ CMS} Collaboration, ``Jet
  Energy Corrections determination at 7 {TeV}'',} \textit{ CMS Physics Analysis
  Summary} \textbf{
  \href{http://cdsweb.cern.ch/record/1308178}{CMS-PAS-JME-10-010}} (2010).

\bibitem{MET}
\href {http://cdsweb.cern.ch/record/1279142} {{ CMS} Collaboration, ``Missing
  Transverse Energy Performance in Minimum-Bias and Jet Events from
  Proton-Proton Collisions at $\sqrt{s}=7$ {TeV}'',} \textit{ CMS Physics
  Analysis Summary} \textbf{
  \href{http://cdsweb.cern.ch/record/1279142}{CMS-PAS-JME-10-004}} (2010).

\bibitem{2010EPJC..tmp..299K}
\hrefCMSnoop {} {{ CMS} Collaboration, ``{CMS} tracking performance results
  from early {LHC} operation'',} \textit{ Eur. Phys. J.} \textbf{ C70} (2010)
  1165, \href{http://www.arXiv.org/abs/1007.1988}{\texttt{ arXiv:1007.1988}}.
  \href{http://dx.doi.org/10.1140/epjc/s10052-010-1491-3}{\texttt{
  doi:10.1140/epjc/s10052-010-1491-3}}.

\bibitem{Maltoni:2002qb}
\hrefCMSnoop {} {F.~Maltoni and T.~Stelzer, ``{MadEvent}: Automatic event
  generation with {MadGraph}'',} \textit{ JHEP} \textbf{ 02} (2003) 027,
  \href{http://www.arXiv.org/abs/hep-ph/0208156}{\texttt{
  arXiv:hep-ph/0208156}}.
\href{http://dx.doi.org/10.1088/1126-6708/2003/02/027}{\texttt{
  doi:10.1088/1126-6708/2003/02/027}}.

\bibitem{PYTHIA}
\hrefCMSnoop {} {T.~Sj{\"o}strand, S.~Mrenna, and P.~Skands, ``{PYTHIA} 6.4
  Physics and Manual'',} \textit{ JHEP} \textbf{ 05} (2006) 026,
  \href{http://www.arXiv.org/abs/hep-ph/0603175}{\texttt{
  arXiv:hep-ph/0603175}}.
  \href{http://dx.doi.org/10.1088/1126-6708/2006/05/026}{\texttt{
  doi:10.1088/1126-6708/2006/05/026}}.

\bibitem{Agostinelli:2002hh}
\hrefCMSnoop {} {{ GEANT4} Collaboration, ``{GEANT4:} A simulation toolkit'',}
  \textit{ Nucl. Instrum. Meth.} \textbf{ A506} (2003) 250.
\href{http://dx.doi.org/10.1016/S0168-9002(03)01368-8}{\texttt{
  doi:10.1016/S0168-9002(03)01368-8}}.

\bibitem{Berger:2009qy}
\hrefCMSnoop {} {E.~L. Berger and Q.-H. Cao, ``Next-to-Leading Order Cross
  Sections for New Heavy Fermion Production at Hadron Colliders'',} \textit{
  Phys. Rev.} \textbf{ D81} (2010) 035006,
  \href{http://www.arXiv.org/abs/0909.3555}{\texttt{ arXiv:0909.3555}}.
\href{http://dx.doi.org/10.1103/PhysRevD.81.035006}{\texttt{
  doi:10.1103/PhysRevD.81.035006}}.

\bibitem{Khachatryan:2010ez}
\hrefCMSnoop {} {{ CMS} Collaboration, ``First Measurement of the Cross Section
  for Top-Quark Pair Production in Proton-Proton Collisions at $\sqrt{s}=7$
  {TeV}'',} \textit{ Phys. Lett.} \textbf{ B695} (2011) 424--443,
  \href{http://www.arXiv.org/abs/1010.5994}{\texttt{ arXiv:1010.5994}}.
\href{http://dx.doi.org/10.1016/j.physletb.2010.11.058}{\texttt{
  doi:10.1016/j.physletb.2010.11.058}}.

\bibitem{Khachatryan:2010xn}
\hrefCMSnoop {} {{ CMS} Collaboration, ``Measurements of Inclusive {W} and {Z}
  Cross Sections in pp Collisions at $\sqrt{s}=7$ {TeV}'',} \textit{ JHEP}
  \textbf{ 01} (2011) 080, \href{http://www.arXiv.org/abs/1012.2466}{\texttt{
  arXiv:1012.2466}}.
\href{http://dx.doi.org/10.1007/JHEP01(2011)080}{\texttt{
  doi:10.1007/JHEP01(2011)080}}.

\bibitem{Campbell:2010ff}
\hrefCMSnoop {} {J.~M. Campbell and R.~K. Ellis, ``{MCFM} for the {Tevatron}
  and the {LHC}'',} \textit{ Nucl. Phys. Proc. Suppl.} \textbf{ 205-206} (2010)
  10, \href{http://www.arXiv.org/abs/1007.3492}{\texttt{ arXiv:1007.3492}}.
\href{http://dx.doi.org/10.1016/j.nuclphysbps.2010.08.011}{\texttt{
  doi:10.1016/j.nuclphysbps.2010.08.011}}.

\bibitem{lumi}
\hrefCMSnoop {} {{ CMS} Collaboration, ``Measurement of {CMS} Luminosity'',}
  \textit{ CMS Physics Analysis Summary} \textbf{
  \href{http://cdsweb.cern.ch/record/1279145}{CMS-PAS-EWK-10-004}} (2010).

\bibitem{Pumplin:2002vw}
\hrefCMSnoop {} {J.~Pumplin {et~al.}, ``New generation of parton distributions
  with uncertainties from global QCD analysis'',} \textit{ JHEP} \textbf{ 07}
  (2002) 012,
\href{http://www.arXiv.org/abs/hep-ph/0201195}{\texttt{ arXiv:hep-ph/0201195}}.

\bibitem{Bayes}
\hrefCMSnoop {} {I.~Bertram {et~al.}, ``A Recipe for the construction of
  confidence limits'',} technical report, 2000.
\newblock
  \href{http://lss.fnal.gov/archive/test-tm/2000/fermilab-tm-2104.pdf}{FERMILAB-TM-2104}.

\end{thebibliography}\endgroup

\cleardoublepage\appendix\section{The CMS Collaboration \label{app:collab}}\begin{sloppypar}\hyphenpenalty=5000\widowpenalty=500\clubpenalty=5000\textbf{Yerevan Physics Institute,  Yerevan,  Armenia}\\*[0pt]
S.~Chatrchyan, V.~Khachatryan, A.M.~Sirunyan, A.~Tumasyan
\vskip\cmsinstskip
\textbf{Institut f\"{u}r Hochenergiephysik der OeAW,  Wien,  Austria}\\*[0pt]
W.~Adam, T.~Bergauer, M.~Dragicevic, J.~Er\"{o}, C.~Fabjan, M.~Friedl, R.~Fr\"{u}hwirth, V.M.~Ghete, J.~Hammer\cmsAuthorMark{1}, S.~H\"{a}nsel, M.~Hoch, N.~H\"{o}rmann, J.~Hrubec, M.~Jeitler, G.~Kasieczka, W.~Kiesenhofer, M.~Krammer, D.~Liko, I.~Mikulec, M.~Pernicka, H.~Rohringer, R.~Sch\"{o}fbeck, J.~Strauss, F.~Teischinger, P.~Wagner, W.~Waltenberger, G.~Walzel, E.~Widl, C.-E.~Wulz
% ...
\vskip\cmsinstskip
% ...
1:~~Also at CERN, European Organization for Nuclear Research, Geneva, Switzerland\\
% ...

\end{sloppypar}
\end{document}